\documentclass{article}
\usepackage{graphicx} 
\usepackage[margin=0.8in]{geometry}  
\usepackage{amsthm}
\usepackage{amsmath}
\usepackage{comment}
\usepackage{amsfonts}
\usepackage{array}
\usepackage{booktabs}
\usepackage{longtable}
\usepackage{hyperref}
\usepackage{multirow}
\usepackage{tikz}
\usepackage{parskip} 
\usepackage[skip=0pt]{caption} 
\usepackage{pdflscape}
\usepackage{graphicx}
\usepackage{authblk}
\usepackage{subcaption}
\usepackage{qcircuit}
\usepackage{subcaption}

\usepackage{amsfonts} 

\usepackage{enumitem}
\usepackage{cite}
\usepackage{setspace}

\normalsize

\title{Generalized Code  Distance through Rotated Logical States in Quantum Error Correction}
\author[1,2]{Valentine Nyirahafashimana}
\author[1,3]{Nurisya Mohd Shah}
\author[4]{ Umair Abdul Halim}
\author[1,5]{Mohamed Othman}
\affil[1]{Institute for Mathematical Research, Universiti Putra Malaysia, 43400 UPM Serdang, Selangor Darul Ehsan, Malaysia}
\affil[2]{Kigali Independent University (ULK), Polytechnic Institute, Kigali Campus, 102 KG 14 Ave Gisozi, Rwanda}
\affil[3]{Department of Physics, Faculty of Science, Universiti Putra Malaysia, 43400 UPM Serdang, Selangor, Malaysia}
\affil[4]{Centre of Foundation Studies in Science, Universiti Putra Malaysia, 43400 UPM Serdang, Selangor, Malaysia}
\affil[5]{Department of Communication Technology and Network, Universiti Putra Malaysia, 43400 UPM Serdang,  Selangor, Malaysia}
\affil[*]{Corresponding author: risya@upm.edu.my; nyirahafashimanav@ulk.ac.rw}

\date{} 

\begin{document}

\maketitle
\begin{abstract}


We construct rotated logical states by applying rotation operators to stabilizer states, extending the logical basis and modifying stabilizer generators. Rotation operators affect the effective code distance $d_R$, which decays exponentially with rotation angles $(\theta, \phi)$, influencing error correction performance.
We quantify the scaling behavior of logical error rates under circuit-level noise, comparing standard depolarizing (SD) and superconducting-inspired (SI) noise models with small and large rotations. Our findings show that the rotated code scales as $0.68d_R (0.65d_R)$ for SD and $0.81d_R (0.77d_R)$ for SI, with small rotation angles leading to a steeper decay of logical error rates. At a physical error rate \((p_{phy})\) of $10^{-4}$, logical errors decrease exponentially with $d_R$, particularly under SI noise, which exhibits stronger suppression. The threshold error rates for rotated logical states are compared with previous results, demonstrating improved resilience against noise. By extending the logical state basis, rotation-based encoding increases error suppression beyond traditional stabilizer codes, offering a promising approach to advancing quantum error correction.
\end{abstract}

\section{Introduction}

In quantum computing, quantum error correction (QEC) and quantum error correction codes (QECC) are essential for the preservation and safeguarding of quantum information, making them fundamental for quantum computation, quantum memories, and quantum communication systems~\cite{steane1998introduction, mummadi2025error,brun2019quantum,devitt2013quantum,forlivesi2024logical,shor1995scheme,ostrev2024classical,zoratti2023improving}. Traditionally, they have been built using the stabilizer formalism, where stabilizer codes are constructed as Abelian subgroups of the Pauli group on \( n \) qubits. While stabilizer codes provide a structured framework for error correction, they must contend with quantum noise arising from environmental interactions and the inherent analogue nature of quantum operations. To ensure reliable computation, fault-tolerant QECCs have been developed to suppress logical errors even when physical errors occur during quantum gates and measurements~\cite{o2024compare,stephens2014fault}. Recent studies show that QECCs achieve high thresholds against depolarizing noise and their performance can be significantly increased by adapting the code to specific noise models, particularly in the presence of structured noise~\cite{bonilla2021xzzx,stephens2013high,tuckett2019tailoring,o2024compare}.

\noindent The surface code~\cite{kitaev2003fault} is a stabilizer code that has one of the highest thresholds of any QECC~\cite{gottesman1997stabilizer}. In their study,~\cite{acharya2024quantum} investigated the effect of increasing the size of a logical qubit on its logical error rate using the surface code. The authors specifically implemented the rotated surface code~\cite{horsman2012surface,bombin2007optimal}, demonstrating that the rotated surface code requires approximately half the number of physical qubits compared to the unrotated surface code~\cite{bravyi1998quantum} while maintaining the same error correction distance. In addition, the threshold scaling of the logical to physical error rates under circuit level noise for both rotated and unrotated codes at high odd and even distances were presented in~\cite{o2024compare}. The authors then compared the number of qubits used by each code to achieve equal logical error rates and found that the rotated code utilizes 74 to 75\% of the number of qubits used by the non-rotated code, depending on the noise model, to achieve a logical error rate of \( p_{log} = 10^{-12} \). A key factor in assessing the effectiveness of such approaches is the error correction distance \(d\), which determines the level of protection against logical errors~\cite{kapshikar2023hardness,valentine2024transforming}. This includes a proved distance formula applicable to codes that exhibit a rotational symmetry $N$ multiple times with respect to the number of qubits and rotational errors~\cite{marinoff2024explicit}. 

\noindent The introduction of a non-commutative Pauli stabilizer formalism has enabled efficient computation of entanglement and local observables, particularly in topological models with non-Abelian anyons that enhance fault tolerance in quantum systems ~\cite{ni2015non,abdollahi2006non}. The extended Clifford group further strengthens this framework by generalizing the Pauli group \(P_n\) to a broader algebraic structure, facilitating advanced quantum gates and error correction techniques~\cite{mastel2023clifford,abdollahi2006non,pillado2013group}. 

\noindent A recent study in $XP$ stabilizer codes establishes an explicit connection between rotation operators and the construction of the logical state. These codes generalized the traditional Pauli stabilizers through a ``quantum lego" approach, allowing efficient tracking of $XP$ symmetries and enabling high-distance error-correcting codes with fault-tolerant properties~\cite{shen2023quantum}. This extension inherently links to rotation-based logical state encoding, where the inclusion of rotation operators in stabilizer codes redefine error syndromes and improve logical error rates. For example, the integration of non-linear QEC (NLQEC), which went beyond strictly linear encoding, provides a framework for extending stabilizer codes under rotation transformations\cite{reichert2022nonlinear}.

\noindent This study aims to extend QEC by constructing rotated logical states and generators using rotation operators \(R_x(\theta)\) and \(R_z(\phi)\) to improve error suppression for complex and non-standard error models. It extends the Pauli and Clifford groups to \(U(2^n)\), incorporating non-Abelian symmetries to broaden logical operations and improve fault tolerance. Additionally, the study explores the effect of rotation operators on code distance and logical error rate scaling under circuit-level noise, considering the role of non-Clifford and logical gates in rotated logical state circuits. By extending stabilizer codes under rotation transformations, this work develops an alternative framework that optimizes QEC performance beyond the conventional stabilizer formalism.
The work is structured as follows: Section~\ref{sec:II} establishes the theories and methods to achieve the generalized rotated logical state. 
Section~\ref{sec:III} analyses the effect of rotation operators on code distance, logical error rate decay, threshold behavior and the advantages of rotation-based error correction.
Section~\ref{sec:IV} concludes with a summary of findings.

\section{Theory and method}\label{sec:II} 
This section explores the non-commutator structure of rotation operators with Pauli operators and extends these transformations to \( U(2^n) \) by incorporating global phase factors and a non-Clifford gate. This expansion introduces new algebraic properties for more flexible encoding beyond stabilizer constraints. The stabilizer state is extended to a rotated logical state, modifying generators and code distance parameters. As the rotation angle increases, the effective code distance decreases, reducing detectable and correctable errors. The physical and logical error rates are quantified using the effective code distance \( d_R \) in the SD and SI noise models to evaluate error correction performance.

\subsection{Non-Commutative Structure in Quantum Systems}

The non-commutation structure refers to the mathematical relationships between operators that do not commute under multiplication. In quantum mechanics and quantum information science, this property is crucial in defining the behavior of quantum systems. Given two operators \( A \) and \( B \), their commuters are defined as:
\begin{equation}
[A, B] = AB - BA.\end{equation}
If the commuter is equal to zero, the commuters commute, meaning that their application order does not affect the outcome. However, if \( [A, B] \neq 0 \), they exhibit a non-commutation structure, leading to fundamental quantum effects. A well-known example of non-commutation appears in quantum computing, is the Pauli matrices \( \sigma_x, \sigma_y, \sigma_z \) satisfy the relation:

\begin{equation}
[\sigma_i, \sigma_j] = 2i \epsilon_{ijk} \sigma_k,   
\label{Levi-civita}
\end{equation}

where \( \epsilon_{ijk} \) is the Levi-Civita symbol, encoding their non-Abelian structure. This structure underlies various quantum operations and error correction mechanisms. In the context of QEC, non-commuting operators influence different types of codes. Like non-stabilizer codes, which include those based on non-Clifford operations, often involve non-commuting generators.

\begin{figure}
    \centering
    \includegraphics[width=0.3\linewidth]{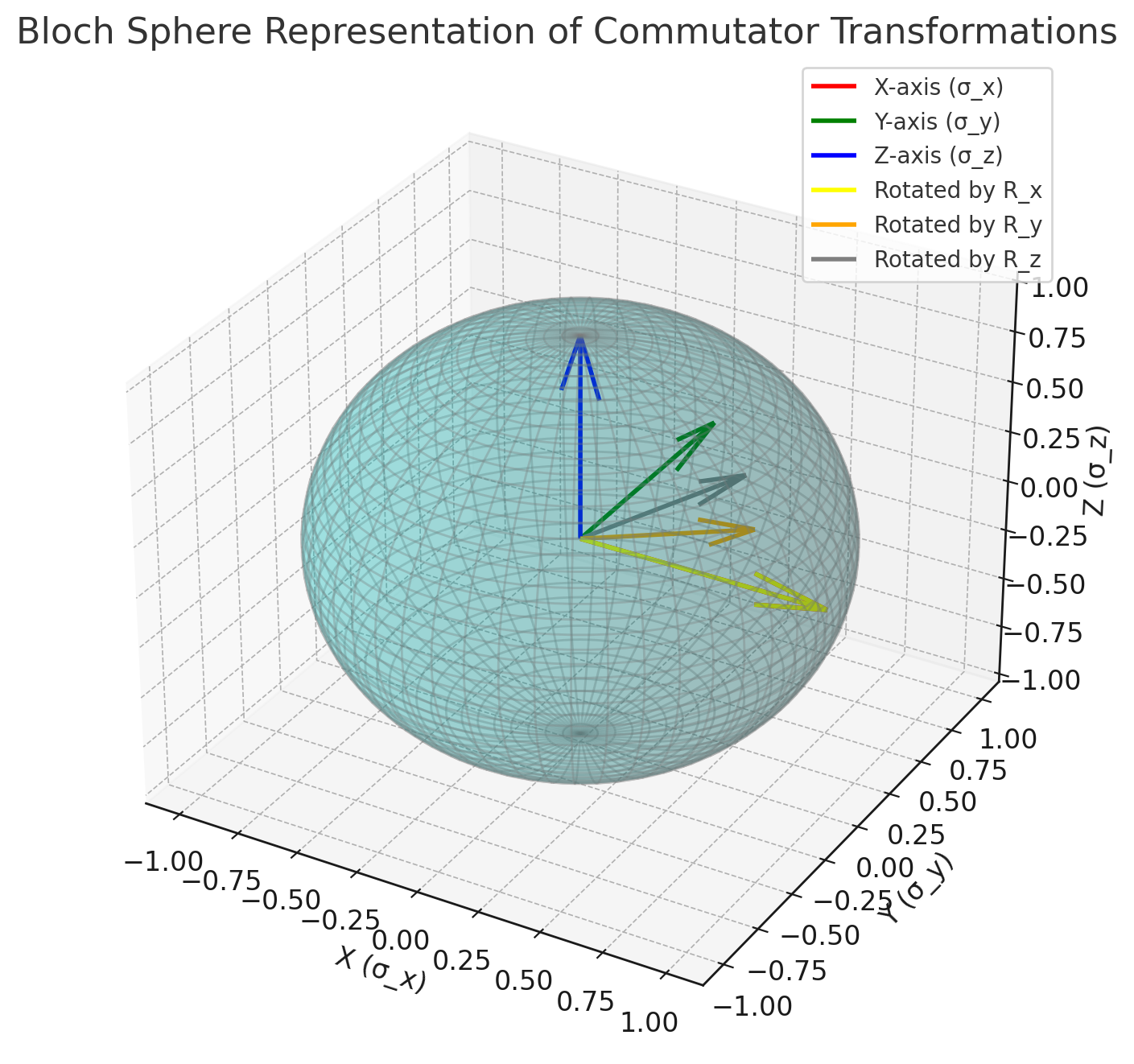}
    \caption{Bloch sphere representation of commutator transformations in~\eqref{matrix} provided from the main non commutator relation in~\eqref{in non-com}}
    \label{fig: Bloch sphere}
\end{figure}

\noindent A group \( G \) is a mathematical structure consisting of elements and a binary operation that satisfies multiplication closure, associativity, identity, and invertibility properties~\cite{john2023group,singh2021mathematical}. The elements of \( G \) also represent quantum operators acting on qubits, some of which belong to non-Abelian groups that do not necessarily commute each other, leading to non-stabilizer groups. Consider \( g, h \in G \), a non-Abelian group satisfies:

\begin{equation}
gh \neq hg.
\label{eq:ineq}
\end{equation} 
Within the code space \( \mathcal{C} \) captures the algebraic structure of \( G \), consisting of states invariant under its action:  
\begin{equation}  
\mathcal{C} = \text{span} \{ | \psi \rangle \in \mathcal{H}^n \mid g | \psi \rangle = | \psi \rangle \, \forall g \in G \}.  
\label{Eq:span}  
\end{equation} 
Non-stabilizer groups extend beyond this, utilizing non-Abelian structures to correct a broader range of errors, inflating QEC capabilities. Extending the stabilizer framework beyond stabilizer formalism involves several strategies, such as incorporation of non-Pauli operators called rotation operators, which maintain orthogonality and commutation relations by satisfying the non-Abelian property with Pauli matrices. These operators generate rotations about their respective Pauli matrix axes in the Bloch-sphere representation. Due to the intrinsic non-commutativity of the Pauli matrices, these rotation operators do not generally commute with the Pauli matrices themselves. The non-commutativity extends to the rotation operators, as expressed by:
\begin{equation}
[R_a(\alpha), \sigma_b] = R_a(\alpha) \sigma_b - \sigma_b R_a(\alpha),
\label{in non-com}
\end{equation}
which, is nonzero unless \( \sigma_b \) is aligned with the rotation axis \( \sigma_a \) .
This non-commutativity leads to a transformation law under conjugation by a rotation operator:
\begin{equation}
R_n(\theta) \sigma_m R_n^\dagger(\theta) = \cos\theta \sigma_m + \sin\theta (\sigma_n \times \sigma_m) + (1 - \cos\theta) (\sigma_n \cdot \sigma_m)\sigma_n.
\label{transf.}
\end{equation}
This approach demonstrates that Pauli matrices do not simply commute with rotations but instead transform according to specific trigonometric mixing rules, dictated by their cross-product and dot-product relationships. The rotation operators, expanded in their trigonometric form,

\begin{equation}
R_a(\alpha) = \cos(\alpha/2) I - i \sin(\alpha/2) \sigma_a,
\label{conjugation}
\end{equation}

further clarify their role in generating continuous transformations on quantum states. They outline the essential non-Abelian nature of spin rotations in quantum mechanics, where state evolution under rotations is governed by the interplay between these operators rather than simple commutation. A rotation operator \( R_a(\alpha) \) commutes with a Pauli operator \( \sigma_b \) when both are aligned along the same axis,\([R_a(\alpha), \sigma_b] = 0\) for \( a = b \). However, if \( a \neq b \), the rotation transforms \( \sigma_b \) into a linear combination of Pauli operators orthogonal to \( a \), resulting in non-commutation \([R_a(\alpha), \sigma_b] \neq 0\) and inducing phase shifts. 
 This is a direct consequence of the non-Abelian nature of Pauli matrices and is fundamental to quantum mechanics and spin dynamics. We evaluate the commutation relation~\eqref{in non-com} by combining \eqref{transf.} and \eqref{conjugation}, the result derived is as follows:
\begin{equation}
[R_a(\alpha), \sigma_b] = (\cos\alpha - 1) \sigma_b + \sin\alpha (\sigma_a \times \sigma_b),
\end{equation}
here \(\alpha\) is rotation angle (\(\theta\) or \(\phi\)) depending on the rotation axis. Then consider \(\alpha\) equal to \(\pi/4\), with \(i\) being the imaginary part,  the numerical results are obtained as:
\begin{equation}
\begin{aligned}
   & [R_x(\theta), \sigma_y]=\begin{bmatrix} 
    0 + 0.7071i & 0 + 0.2929i \\ 
    0 - 0.2929i & 0 - 0.7071i 
    \end{bmatrix},\quad  [R_x(\theta), \sigma_z]=\begin{bmatrix} 
    -0.2929 + 0i & -0.7071 + 0i \\ 
    0.7071 + 0i & 0.2929 + 0i 
    \end{bmatrix} \\
    & R_y(\theta), \sigma_x]=\begin{bmatrix} 
    0 - 0.7071i & -0.2929 + 0i \\ 
    -0.2929 + 0i & 0 + 0.7071i \end{bmatrix}, \quad [R_y(\theta), \sigma_z]= \begin{bmatrix} 
    -0.2929 + 0i & 0 + 0.7071i \\ 
    0 + 0.7071i & 0.2929 + 0i 
    \end{bmatrix}\\
    & R_z(\phi), \sigma_x]=\begin{bmatrix} 
    0 + 0i & 0.4142 + 0i \\ 
    -1 + 0i & 0 + 0i 
    \end{bmatrix}, \quad [R_z(\phi), \sigma_y]=\begin{bmatrix} 
    0 + 0i & 0 - 0.4142i \\ 
    0 - 1i & 0 + 0i .
    \end{bmatrix}
\end{aligned}
\label{matrix}
\end{equation}
 Equation~\eqref{matrix} behave as expected, indicating the non-commutativity and the non-Abelian nature of rotation operators and Pauli matrices in quantum mechanics. Fig.~\ref{fig: Bloch sphere} visualizes Bloch sphere representation of the commutation transformations. The initial Bloch vector (aligned along the x-axis) is transformed by the rotation matrices \(R_x, R_y\) and \(R_z\), shifting its position on the sphere.

\noindent Generally, a rotation operator \( R_{\hat{n}}(\theta) \), defined for a rotation angle \(\theta\) about an axis \(\hat{n} = (n_x, n_y, n_z)\), is expressed as:  

\begin{equation}
R_{\hat{n}}(\theta) = e^{-i \frac{\theta}{2} (\hat{n} \cdot \boldsymbol{\sigma})},
\label{general op}
\end{equation}
where \(\boldsymbol{\sigma} = (\sigma_x, \sigma_y, \sigma_z)\) are the Pauli matrices and \(\hat{n} \cdot \boldsymbol{\sigma} = n_x \sigma_x + n_y \sigma_y + n_z \sigma_z\), represents their projection along the axis. The transformation of a Pauli matrix \(\sigma_j\) by a rotation operator depends on their commutation, which is determined using the Baker-Campbell-Hausdorff formula~\cite{van2015special}.  
\begin{equation}
R_{\hat{n}}(\theta) \sigma_j R_{\hat{n}}^\dagger(\theta) = e^{-i \frac{\theta}{2} (\hat{n} \cdot \boldsymbol{\sigma})} \sigma_j e^{i \frac{\theta}{2} (\hat{n} \cdot \boldsymbol{\sigma})}.
\end{equation}
If \(\sigma_j\) is parallel to \(\hat{n}\), the operator commutes, leaving \(\sigma_j\) unchanged. If not, the operator introduces rotation to the basis. A rotation about the \(z\)-axis (\(\hat{n} = (0, 0, 1)\)) with \(\sigma_x\) gives:  
\[
R_{\hat{z}}(\theta) = e^{-i \frac{\theta}{2} \sigma_z}, \quad \text{and the transformation:} \quad R_{\hat{z}}(\theta) \sigma_x R_{\hat{z}}^\dagger(\theta) = \cos(\theta) \sigma_x + \sin(\theta) \sigma_y.
\]
This demonstrates that  \(R_{\hat{n}}(\theta)\) does not commute with \(\sigma_j\), instead, rotating it into a combination of \(\sigma_x\) and \(\sigma_y\). Conjugation by rotation gates in the Pauli basis are given by:
\begin{equation}
R_{\boldsymbol{\sigma}}(\theta) \sigma_j R_{\boldsymbol{\sigma}}(-\theta) = \sigma_j \cos(2\theta) + (\boldsymbol{\sigma} \times \sigma_j) \sin(2\theta) + (\boldsymbol{\sigma} \cdot \sigma_j) \boldsymbol{\sigma} (1 - \cos(2\theta)),  \label{conjugation 1}
\end{equation}
where \( \boldsymbol{\sigma} \) represents the Pauli vector. Under this transformation, any Pauli operator \( \sigma_j \) transforms accordingly. Rotation operators, as non-Pauli operations, extend beyond individual Pauli matrices and play a crucial role in quantum transformations. They describe qubit rotations on the Bloch sphere (see Fig.\ref{fig: Bloch sphere}), illustrating how \( x \)-errors mix into \( y \)-errors and how errors propagate through gates, impacting error detection and correction. These operators also define logical operations in non-stabilizer codes, enabling quantum information manipulation beyond the stabilizer formalism. Moreover, they contribute to designing error-resilient rotations for fault-tolerant quantum computing and facilitate logical operator mapping in rotated surface codes~\cite{o2024compare}.

\subsection{Extension of Pauli and Clifford groups}
\subsubsection*{\(U(2^n)\)}

  The Pauli and Clifford groups include non-commutative elements, thereby broadening the set of logical operations and significantly boosting fault tolerance. In \(U(2^n)\), the extended Pauli group for \( n \) qubits, denoted as \( G^{\text{ext}}_{\text{Pauli}} \), consists of all possible tensor products of Pauli matrices acting on \( n \) qubits, along with global phase factors \( \{ \pm 1, \pm i \} \). Mathematically, it is defined as 
\begin{equation}
G^{ext}_{\text{Pauli}} = \{ \pm 1, \pm i \} \cdot \langle I, X, Y, Z \rangle^{\otimes n},
\label{Pauli G}
\end{equation} 

For a single qubit, the Pauli group \( \mathcal{P}_1 \) consists of 16 elements, given by \( \mathcal{P}_1 = \{ \pm I, \pm X, \pm Y, \pm Z, \pm iI, \pm iX, \pm iY, \pm iZ \} \), while, the \( n \)-qubit \( \mathcal{P}_n \) consists of all possible tensor products of these single-qubit \( \mathcal{P}_1 \), leading to \( 4^n \) distinct Pauli operators. This extension further increases the total number of elements by a factor of 4 from the global phase factors. Each element of \( G^{\text{ext}}_{\text{Pauli}} \) takes the form:
\begin{equation}
    P_n=\{\lambda P_1\otimes P_2 \dots \otimes P_n |P_i \in \{I,X,Y,Z\},\lambda \in\{\pm 1,\pm i\}\}
\end{equation}
Thus, the order of \( G^{\text{ext}}_{\text{Pauli}} \) for \( n \) qubits is given by the formula:  

\begin{equation}
|G^{\text{ext}}_{\text{Pauli}, n}| = 2^{2n+2}, \quad \text{which grows exponentially with \( n \)}
\end{equation}
It is a non-Abelian structure due to the non-commutative nature of Pauli matrices. The Pauli group without phase factors forms a normal subgroup within it.
The extension of single-qubit Pauli matrices to multi-qubit systems through tensor products and phase factors, its order grows exponentially with the number of qubits, making it a fundamental tool in quantum computing and information theory.

\noindent  Whereas, the Clifford group \( \mathcal{C} \), given by 
\begin{equation}
\mathcal{C} = \{ U \in \mathcal{U}(2^n) \mid U G^{ext}_{\text{Pauli}} U^\dagger = G^{ext}_{\text{Pauli}} \},\label{Clifford C}
\end{equation} 
that normalizes \( G^{ext}_{\text{Pauli}} \) in the \( n \)-qubit unitary group \( \mathcal{U}(2^n) \), meaning it maps Pauli operators onto themselves under conjugation. The extended Clifford group denoted as \( \mathcal{C}_{\text{extended}} \) is generated by the Clifford group \( \mathcal{C} \) and an additional non-Clifford gate \( \mathfrak{G} \), to support error correction and logical operations. This can be expressed as :
\begin{equation}
\mathcal{C}_{\text{extended}} = \langle \mathcal{C}, \mathfrak{G} \rangle, \implies \mathcal{C}_{\text{extended}} = \{ U \in \mathcal{U}(2^n) \mid U \cdot  G^{ext}_{\text{Pauli}} \cdot U^\dagger \subseteq \langle G^{ext}_{\text{Pauli}}, \mathfrak{G} \rangle \},
\label{C_ext}
\end{equation} 
where the added gates expand the group’s capabilities as: \begin{equation}
\mathfrak{G}=\{\text{T(\(\pi/8\)), Controlled-Hadamard(CH), Toffoli(CCNOT), Controlled-S (CS) gate ,\(\dots\)}\}, \label{non-Clifford gate}\end{equation}
that ensures the compatibility between \( G^{ext}_{\text{Pauli}} \) and the extended set of logical operations.
It consists of unitary operators \( U \in \mathcal{U}(2^n) \) that transform \( G^{ext}_{\text{Pauli}} \) containing stabilizer and non-stabilizer codes into a subset of the group \( \langle G^{ext}_{\text{Pauli}}, \mathfrak{G} \rangle \), which is generated by combining \( G^{ext}_{\text{Pauli}} \) with \( \mathfrak{G} \).
\(|\mathcal{C}_n|\) order is approximately given by \(|\mathcal{C}_n| \approx 2^{n^2 + 2n}\), reflecting the number of possible transformations that preserve the Pauli structure. However, extending the Clifford group by including additional non-Clifford gates, such as the T gate, leads to \(\mathcal{C}_{\text{extended}, n}\), which has a larger order of
\(|\mathcal{C}_{\text{extended}, n}| \approx 2^{n^2 + 2n + 1}\). 
\noindent The \( G^{ext}_{\text{Pauli}} \) and \( \mathcal{C}_{\text{extended}}\) satisfy the non-commutation relations:

\begin{equation}
P_1 P_2 = \lambda P_2 P_1, \quad \lambda \in \{\pm 1, \pm i\},
\end{equation} 
where \( P_1, P_2 \in G^{ext}_{\text{Pauli}} \). The approach ensures that unitary transformations \( U \) preserve the non-commutation relations for extended Pauli group under conjugation:
\begin{equation}
U P_1 U^\dagger \cdot U P_2 U^\dagger = \lambda \, U P_2 U^\dagger \cdot U P_1 U^\dagger.
\end{equation}
On the side of extended Clifford group \( \langle G^{ext}_{\text{Pauli}}, \mathfrak{G} \rangle \) is further enriched by elements \( P_i \in G^{ext}_{\text{Pauli}} \) and \( g_i \in \mathfrak{G} \), combined as:
\begin{equation}
Q = P_1 g_1 P_2 g_2 \cdots P_k g_k, \quad\text{where the closure condition requires } \quad U P U^\dagger = \sum_j c_j Q_j, \quad Q_j \in \langle G^{ext}_{\text{Pauli}}, \mathfrak{G} \rangle.
\end{equation}
The non-Clifford gates inflates the versatility of \( \mathcal{C}_{\text{extended}} \), allowing universal quantum computation and expanded error-correction capabilities. Thus, the non-commutation relationship is a fundamental property of \(G^{ext}_{\text{Pauli}}\), reflecting the inherent non-Abelian structure essential to encode and manipulate quantum information.

\noindent Unlike the standard Pauli group, which is confined to stabilizer-based error correction, the \(G^{ext}_{\text{Pauli}}\) introduces additional algebraic properties that permit more flexible encoding schemes beyond stabilizer constraints. Building on this, \(\mathcal{C}_{\text{extended}}\) integrates extra gates, including non-Clifford elements, granting access to non-stabilizer resources such as magic states and non-Gaussian operations. This extension reinforces QEC by broadening the available computational and fault-tolerant strategies.  
Despite these advancements, both the \(G^{ext}_{\text{Pauli}}\) and \(\mathcal{C}_{\text{extended}}\) remain fundamentally limited within \( U(2^n) \). The \(G^{ext}_{\text{Pauli}}\) lacks universality, cannot perform entangling operations, and is restricted to discrete transformations. Similarly, while \(\mathcal{C}_{\text{extended}}\) normalizes the Pauli group and expands computational power, it still cannot approximate arbitrary unitary operations, preventing it from achieving universal quantum computation.  
To overcome these constraints, \( \mathfrak{G} \) and arbitrary rotations gate are required. These gates break the normalizer condition, granting access to the full unitary space and enabling universal quantum computation. Their inclusion is crucial for fault-tolerant logical operations, bridging the gap between structured error correction and complete quantum universality~\cite{leone2023clifford}.

\subsection{Construction of Rotated Logical States under rotation operators}

This part extends stabilizer logical states to rotated logical states and their generators under rotation operators like \( R_x(\theta) \) and \( R_z(\phi) \). It quantifies the physical (\( p_{phy} \)) and logical (\( p_{log} \)) error rates to assess code's efficiency. For larger X- and Z-direction rotations, the approach incorporates non-Pauli rotations into stabilizer codes, optimizing error correction and enabling flexible encoding to handle standard depolarization (SD) and superconductor-inspired (SI) noise models effectively~\cite{bausch2023learning}.

\subsubsection{Generalized Code Space with Rotation Operators}

Let \( \mathcal{H} \) be the Hilbert space of dimension \( 2^n \) of \( n \) qubits system. We define a generalized code space \( \mathcal{C}_R \) that extends the stabilizer formalism beyond it by taking rotation operators into the unitary group \( \mathcal{G}_R \), such that:

\begin{equation}
\mathcal{C}_R = \text{span} \left\{ |\psi_j\rangle : |\psi_j\rangle = U_R |\phi_j\rangle, U_R \in \mathcal{G}_R, |\phi_j\rangle \in \mathcal{H}_0 \right\},
\end{equation}

where \( \mathcal{H}_0 \subset \mathcal{H} \) is a reference subspace, and \( \mathcal{G}_R \) includes both Clifford operations and continuous rotations, meaning that \( U_R \) can be expressed as:

\begin{equation}
U_R = U_C R_{\hat{n}}(\alpha),
\end{equation}

with \( U_C \in \mathcal{C} \) and \(R_{\hat{n}}(\alpha)\) being a rotation operator of \( SU(2^n) \).
 \( \mathcal{C}_R \) is a valid subspace of \( \mathcal{H} \) , as proved and verified in  its closure under linear combinations, inner product preservation, and unitary operations~\cite{cariolaro2015vector}.
First, \( \mathcal{C}_R \) is spanned by transformed basis states \( |\psi_j\rangle \), meaning any linear combination remains in \( \mathcal{C}_R \):

\begin{equation}
\sum_j \alpha_j |\psi_j\rangle = \sum_j \alpha_j U_R |\phi_j\rangle, \quad U_R \in  \mathcal{C}_R.
\end{equation}
Since rotation operators are linear, we obtain:

\begin{equation}
U_R \left(\sum_j \alpha_j |\phi_j\rangle \right) = \sum_j \alpha_j U_R |\phi_j\rangle,
\end{equation}

showing that \( \mathcal{C}_R \) is closed under superposition, \( \alpha_i\) is complex coefficients that determine the superposition of the quantum states. Next, inner products are preserved due to the unitarity of \( U_R \), satisfies the following:

\begin{equation}
\langle \psi_i | \psi_j \rangle = \langle \phi_i | U_R^\dagger U_R | \phi_j \rangle = \langle \phi_i | \phi_j \rangle.
\end{equation}
Thus, \( \mathcal{C}_R \) retains the orthogonality structure of \( \mathcal{H}_0 \).
The dimensionality of \( \mathcal{C}_R \) expands beyond traditional stabilizer codes, if \( \mathcal{H}_0 \) has dimension \( k \), and \( \mathcal{G}_R \) generates a continuous set of transformations, then:

\begin{equation}
\dim(\mathcal{C}_R) \geq k.
\end{equation}

This implies that \( \mathcal{C}_R \) is a richer structure compared to discrete stabilizer codes.
Lastly, closure under unitary transformations is verified for \( |\psi_j\rangle \in \mathcal{C}_R \), then applying another unitary \( V_R \in \mathcal{G}_R \) results in:

\begin{equation}
V_R |\psi_j\rangle = V_R U_R |\phi_j\rangle.
\end{equation}

Since \( V_R U_R \in \mathcal{G}_R \), the transformed state remains in \( \mathcal{C}_R \), ensuring its closure under unitary operations. In this approach of introducing rotation operators in the transformation set \( \mathcal{G}_R \), we extend the stabilizer framework beyond its conventional discrete structure. The \( \mathcal{C}_R \) retains key stabilizer properties while allowing continuous transformations which includes rotation operators, making it useful for fault-tolerant quantum computation, logical gate deformations, and hybrid quantum error correction schemes.

\subsubsection{Definition of stabilizer state}

A stabilizer state \( |\psi\rangle \) is a unique quantum state that remains invariant under an Abelian subgroup \( \mathcal{S}_{\psi} \) of the \( n \)-qubit Pauli group \( P_n \), generated by \( n \) independent elements. This subgroup, called the stabilizer, excludes \(-1\) and satisfies \( \mathcal{S}_{\psi} \subseteq U_{\psi} \subseteq G_{\psi} \), where \(\mathcal{S}\) is stabilizer group, \(U\) represent normalizer group that includes all Pauli operators, and \(G\) is symmetry group~\cite{englbrecht2020symmetries}.
Let \(\mathcal{H} = (\mathbb{C}^2)^{\otimes n}\) be the Hilbert space of an \(n\)-qubit quantum system. A stabilizer state \( |\psi\rangle \)~\cite{bombin2012universal}, is a pure state satisfying the condition:
\begin{equation}
\mathcal{C} = \{ |\psi\rangle : S_j |\psi\rangle = |\psi\rangle, \forall S_j \in \mathcal{S} \}.
\end{equation}

For a general \( n \)-qubit bit-flip code, the logical qubits \( |0_L\rangle \) and \( |1_L\rangle \) are encoded using \( n \) physical qubits to protect against single-qubit bit-flip errors. The stabilizer group \( \mathcal{S} \), generated by \( S_j = Z_j Z_{j+1} \) for \( j= 1, 2, \dots, n-1 \), ensures the invariance of the code space. The stabilizer logical basis states can be defined as:

\begin{equation}
|0_L\rangle = \frac{1}{\sqrt{2^{n-1}}} \sum_{\mathbf{v} \in \mathcal{V}} |\mathbf{v}\rangle, \quad |1_L\rangle = \frac{1}{\sqrt{2^{n-1}}} \sum_{\mathbf{v} \in \mathcal{V}} |\mathbf{v} \oplus \mathbf{1}\rangle,
\label{stab0}
\end{equation}

where \( \mathcal{V} \) is the set of binary strings of length \( n \), with all bits 0 or 1 (that is, \( \mathcal{V} = \{0^n, 1^n\} \)), \( \mathbf{1} = (1, 1, \dots, 1) \) is the all-ones vector and \( \oplus \) denotes the bitwise XOR operation. Specifically, the developed approach provides the following.

\begin{equation}
|0_L\rangle = \frac{1}{\sqrt{2^{n-1}}}(|0^n\rangle + |1^n\rangle), \quad |1_L\rangle = \frac{1}{\sqrt{2^{n-1}}}(|1^n\rangle + |0^n\rangle).
\label{stab1}
\end{equation} 
The logical state \( |\psi\rangle \) in the stabilizer code space expressed as a superposition of \( |0_L\rangle \) and \( |1_L\rangle \), is obtained as:  
\begin{equation}
|\psi\rangle = \alpha |0_L\rangle + \beta |1_L\rangle, \quad |\psi\rangle=\frac{\alpha +\beta}{\sqrt{2^{n-1}}}(|0^n\rangle+\beta |1^n\rangle).
\label{stab logical state}
\end{equation} 

\subsubsection{Applying Rotation Operators to Stabilizers state}
To extend this framework beyond the stabilizer formalism, 
we apply rotation operators such as \( R_x(\theta) \) and \( R_z(\phi) \) to ~\eqref{stab logical state}, which are continuous rotations about the \( X \)- and \( Z \)-axes of the Bloch sphere. These rotation operators are defined as:
\begin{equation}
R_x(\theta) = e^{-i\theta X/2} = \cos(\theta/2)I - i\sin(\theta/2)X, \quad R_z(\phi) = e^{-i\phi Z/2} = \cos(\phi/2)I - i\sin(\phi/2)Z.
\label{rotation operators}
\end{equation}
Here, \(I, X, Z\) are Pauli operators, \(\theta\) and \(\phi\) are the rotation angle with respect to the axis. For an \( n \)-qubit system, the rotation operator extends to the \( n \)-qubit space as a tensor product of \( 2 \times 2 \) rotation matrices.
\begin{equation}
R_x(\theta)^{\otimes n} = R_x(\theta) \otimes R_x(\theta) \otimes \cdots \otimes R_x(\theta), \quad
R_z(\phi)^{\otimes n} = R_z(\phi) \otimes R_z(\phi) \otimes \cdots \otimes R_z(\phi) \end{equation}
 When applied, these operators modify the stabilizers, leading in the transformations law given in~\eqref{transf.}, we obtain:
\begin{equation}  
R_x(\theta) Z R_x(\theta)^\dagger = \cos(\theta) Z - i \sin(\theta) Y , \quad  R_z(\phi) X R_z(\phi)^\dagger = \cos(\phi) X + i \sin(\phi) Y .\end{equation}

The stabilizer logical state encoding scheme is modified with rotation operators to the logical basis states in~\eqref{stab1}, transforming them into non-stabilizer basis states. This transformation is performed using \( R_x(\theta)^{\otimes n} \) and \( R_z(\phi)^{\otimes n} \) to n-qubit system, which act independently on each qubit. As a result, the original stabilizer logical basis states are redefined, producing rotated logical basis states, expressed as:

\begin{equation}
|0^R_L\rangle = R_x(\theta)^{\otimes n} R_z(\phi)^{\otimes n} |0_L\rangle, \quad |1^R_L\rangle = R_x(\theta)^{\otimes n} R_z(\phi)^{\otimes n} |1_L\rangle.
\label{begata}
\end{equation}  

Substituting \(|0_L\rangle\) and \(|1_L\rangle\), the transformed states become:  

\begin{equation}
|0^R_L\rangle = \frac{1}{\sqrt{2^{n-1}}} R_x(\theta)^{\otimes n} R_z(\phi)^{\otimes n} \left(|0^n\rangle + |1^n\rangle\right), \quad
|1^R_L\rangle = \frac{1}{\sqrt{2^{n-1}}} R_x(\theta)^{\otimes n} R_z(\phi)^{\otimes n} \left(|1^n\rangle + |0^n\rangle\right).
\label{non-stab1}
\end{equation} 
 The rotation operator acts individually on qubit \(|\psi\rangle\) in \eqref{stab logical state}, leading to: 
\[R_x(\theta)^{\otimes n} R_z(\phi)^{\otimes n} |\psi\rangle = \alpha R_x(\theta)^{\otimes n} R_z(\phi)^{\otimes n} |0_L\rangle + \beta R_x(\theta)^{\otimes n} R_z(\phi)^{\otimes n} |1_L\rangle. \]

Therefore, rotated logical state \(|\psi_R \rangle\) with independent qubit-wise rotations to the logical qubit state \(|\psi\rangle\) is expressed as:
\begin{equation}
|\psi_R\rangle = R_x(\theta)^{\otimes n} R_z(\phi)^{\otimes n} |\psi\rangle.
\end{equation}
Using the definitions in~\eqref{stab0} and simplifying, the new rotated logical state becomes:
\begin{equation}
|\psi_R\rangle = \frac{\alpha + \beta}{\sqrt{2^{n-1}}} \left( |R_0\rangle^{\otimes n} + |R_1\rangle^{\otimes n} \right),
\label{new state}
\end{equation}
where, 
\[|R_0\rangle = R_x(\theta) R_z(\phi) |0\rangle, \quad |R_1\rangle = R_x(\theta) R_z(\phi) |1\rangle.\]

The rotated logical state remains a superposition of transformed computational basis states. Each qubit undergoes the same single qubit transformation \( R_x(\theta) R_z(\phi) \). The logical encoding structure is preserved but is now generalized to include continuous transformations.
This state extends beyond discrete operations by incorporating continuous logical transformations, introducing continuous degrees of freedom and also  a logical encoding of rotation gates, which are useful for fault-tolerant quantum computation.

\subsubsection{Modified stabilizer generators }

In an \( n \)-qubit system, the stabilizer generators \( S_j \) define a group that preserves the logical subspace. This generator \( S_j \) is typically expressed as a tensor product of Pauli matrices acting on different qubits~\cite{mello2024hybrid,yoshida2010framework}. For a one-dimensional chain of qubits, a standard stabilizer generator can be written as:

\begin{equation}
S_j = Z_j Z_{j+1}.
\label{sin gen}
\end{equation}
More generally, the stabilizer group is generated by operators of the form:
\begin{equation}
S_j = \bigotimes_{i=1}^n P_i,
\label{gen gen }
\end{equation}

where each \( P_i \) is a Pauli matrix (\( I, X, Y, Z \)) acting on the \( i^{th}\)-qubit. When rotation operators are applied to~\eqref{sin gen}, become rotated generators \( S^R_j \), which no longer commute as the original stabilizer group.
Now, the single-qubit system under rotation operators to the entire system becomes:

\begin{equation}
\begin{aligned}
&S^R_j = R_x(\theta)^{\otimes n} R_z(\phi)^{\otimes n} S_j R_z(-\phi)^{\otimes n} R_x(-\theta)^{\otimes n},\\
&S^R_j = \cos^2(2\theta) Z_j Z_{j+1} - \cos(2\theta) \sin(2\theta) (Z_j Y_{j+1} + Y_j Z_{j+1}) + \sin^2(2\theta) Y_j Y_{j+1}.
\end{aligned}
\label{eq gen}
\end{equation}
Each stabilizer \( S_j \) is conjugated by the rotation operators, modifying it into a rotated generator \( S^R_j \). Equation~\eqref{eq gen} is performed under the transformation rules for individual Pauli operator for rotation about the X- and Z- axes in~\eqref{conjugation 1} and rotated generator under rotation for each Pauli matrix \( P_i \) transforms according to it, leading to:

\begin{equation}
S^R_j = \sum_k c_k \bigotimes_{i=1}^n P_i^{(k)},
\label{new generators}
\end{equation}

where \( c_k \) represents coefficients that depend on the rotation angles, and each \( P_i^{(k)} \) is a transformed Pauli operator. 
The rotated generator \( S^R_j \) is no longer purely Pauli operators, but a linear combination of multiple Pauli terms, meaning that they do not commute as in the original stabilizer group. This introduces continuous deformations of stabilizer codes, extending them beyond discrete Clifford operations. The resulting non-stabilizer generators allow logical gate deformations crucial for fault-tolerant quantum computation and are useful for logical qubit rotations and error correction under continuous transformations.
\subsubsection{Modification of Code Parameters }

In a stabilizer code, the parameters \([[n, k, d]]\) define its structure and error-correcting capability, where \(n\) is the number of physical qubits, \(k\) the logical qubits, and \(d\) the code distance, indicating error tolerance. While rotations preserve \( n \) and \( k \), but alter \( d \) by transforming stabilizer generators into non-Pauli forms. This continuous deformation affects error detection and correction properties. Since the stabilizer group still spans a subspace of \( 2^{n-k}\) dimensions, \( n \) and \( k \) remain unchanged. However, the code distance \( d \), which is the minimum weight of an operator that maps one logical codeword to another, can be affected. Logical operators transform under rotation operators as:
\begin{equation}
X_L^R = R_x(\theta)^{\otimes n} R_z(\phi)^{\otimes n} X_L R_z(-\phi)^{\otimes n} R_x(-\theta)^{\otimes n}, \quad Z_L^R = R_z(\phi)^{\otimes n} R_x(\theta)^{\otimes n} Z_L R_x(-\theta)^{\otimes n} R_z(-\phi)^{\otimes n},
\label{logical operators}\end{equation} 
since \(X_L^R\) now includes linear combinations of Pauli terms. To account for the impact of non-Pauli terms on error correction, effective code distance \( d_R \) is defined as:

\begin{equation}
d_R = d \cdot f(\theta, \phi),
\end{equation}

where \( f(\theta, \phi) \) is a deformation function that depends on the rotation angles \( \theta \) and \( \phi \) which quantifies the reduction in error correction capacity. A possible form of \( f(\theta, \phi) \), based on the perturbative analysis of stabilizer deformations, is

\begin{equation}
f(\theta, \phi) = e^{-\lambda (\theta^2 + \phi^2)},
\end{equation}

here, \( \lambda \) is a decoherence parameter that depends on the stabilizer structure, and noise model~\cite{bausch2023learning,bausch2024learning}, the exponential decay function indicates that as the magnitude of rotations \((\theta,\phi)\) increases, the fidelity of the encoded state decreases. A larger \(\lambda\) implies stronger decoherence, meaning that small deviations in \(\theta\) and \(\phi\)  lead to a rapid loss of stabilizer quantum information. Thus, the effective code distance is given by

\begin{equation}
d_R = d  e^{-\lambda (\theta^2 + \phi^2)}. \label{dR}
\end{equation}

This approach implies that for small rotation angles (\(\theta, \phi \approx 0\)), \( f(\theta, \phi) \approx 1 \), which means \( d_R \approx d \), the error-correcting capability remains intact. However, for large rotations \((\theta ,\phi\gg 0)\), where the stabilizers mix significantly in non-Pauli terms, \( f(\theta, \phi) \) decreases, leading to \( d_R < d \), effectively reducing the number of detectable and correctable errors while increasing the ability to address errors beyond stabilizer scheme.

\subsection{Quantifying Physical and Logical Error Rates for the Rotated Logical State}

In QEC, the relationship between the \textit{physical error rate} ( \( p_{\text{phy}} \)) and the \textit{logical error rate} (\( p_{\text{log}} \)) determines the effectiveness of an error-correcting code. For a stabilizer code with parameters \( [[n, k, d]] \), the logical error rate follows a power-law scaling~\cite{o2024compare}, where the \textit{code distance} (\( d \)) governs the suppression of logical errors. When a stabilizer code undergoes rotated logical qubit state, the effective code distance \((d_R)\) is reduced as in~\eqref{dR}, where \( \lambda \) is a decoherence parameter capturing the sensitivity of the code distance to rotation angles \( \theta \) and \( \phi \). Consequently, the \( p_{\text{log}}^R \) rotated is modified as 

\begin{equation}
p_{\text{log}}^R \approx A p_{\text{phy}}^{(d_R+1)/2},
\label{log error}
\end{equation}
increasing for large rotations due to the reduction in \( d_R \). A crucial aspect of error correction is the \textit{threshold error rate} (\( p_{\text{th}} \)), which defines the maximum tolerable \( p_{\text{phy}}\) before logical errors become uncontrollable. Under rotation, this threshold is suppressed with \(d_R\)
making the system more vulnerable to errors for large rotations. For sufficiently large \( d_R \), the \( p_{\text{log}}^R \) is expected to follow the relation:  

\begin{equation}  
p_{\text{log}}^R  = (p_{\text{phy}} - p_{\text{th}}) d_R^{1/\nu_0}   \label{thre-lo-phy}
\end{equation}  
where \( \nu_0 \) denotes the scaling exponent associated with the universality class of the SD and SI noise model~\cite{stephens2014fault}. 
However, SI noise, \(p_{\text{th}}\) is typically lower than that of depolarizing noise due to correlated noise sources and leakage errors.
\begin{equation}
     p_{\text{th}}^R = p_{\text{th}} d_R^{-1/\nu_0}.
    \label{eq:thres SI}
\end{equation}

A higher \( p_{\text{th}} \) is desirable for QECC performance, while a lower logical error probability ensures effective error suppression. Below the threshold, \( p_{\text{log}}\) scales with \( p_{\text{phy}}\) following a power law, reflecting exponential error suppression with increasing code distance~\cite{o2024compare}. The logical error probability \( p_{\text{log}} \) can be expressed as:  
\begin{equation}
   p^R_{\text{log}} = \sum_{i}^{n} \alpha_i \left(\frac{p_{\text{phy}}}{p_{\text{th}_i}}\right)^{d_R + i} \label{equation.Plog}
\end{equation}
where the \(d_R\) is given in~\eqref{dR}, with  \( d = d/2 \) for even code distances, and \( d = (d+1)/2 \) for odd code distances, and \(n\) represents the number of qubit in the system~\cite{o2024compare,dennis2002topological,fowler2012surface} . For a large \(d_R\), the dominant term follows a similar scaling to the ~\eqref{thre-lo-phy}, here the leading order behavior is: 
\begin{equation}   
p^R_{\text{log}} \sim \left(\frac{p_{\text{phy}}}{p_{\text{th}}}\right)^{d_R}\end{equation}
The impact of these rotations is further analyzed under two noise models: SD noise, where qubits experience independent Pauli errors, and SI noise, where dephasing (\( Z \)-bias) dominates~\cite{o2024compare,stephens2014fault}. In SD noise, rotations introduce non-Pauli errors, increasing the effective \(p_{\text{phy}}\), leading to a faster degradation of error correction. 
In contrast, SI noise initially exhibits better error scaling, but still suffers from rotation-induced loss of bias preservation. In this case, the \(p^R_{\text{log}}\) follows a different scaling, power-law relation with (\(d_R\)) in exponent term of the SD and SI noise models, respectively: \((d_R+1)/2\) and \((d_R+2)/2\). SI noise has built-in bias-preserving properties and logical errors scale more favorably than in the SD noise case.

\noindent In terms of higher order, \eqref{equation.Plog} failed to capture the scaling relationship of all data in n-qubits system. Then, we generalize the effect of rotations on error scaling, by introducing a fitted function of the form:
\begin{equation}
p_{\text{log}}^R = A p_{\text{phy}}^{\xi(\theta, \phi)},
\label{fitted function}
\end{equation}
where \( \xi(\theta, \phi) \) is the modified exponent that incorporates the effective code distance \( d_R \). Comparing  ~\eqref{fitted function}  with ~\eqref{log error} to identify  \( \xi(\theta, \phi) \) for the SI and SD noise models, we obtain:
\[\xi^{\text{SI}}(\theta, \phi) = \frac{d_R + 2}{2}, \quad \xi^{\text{SD}}(\theta, \phi) = \frac{d_R + 1}{2}. \] 
The standard scaling form used in QEC~\cite{o2024compare} is given:
\begin{equation*}
p_{\text{log}} = \alpha \left( \frac{p_{\text{phy}}}{\beta} \right)^{\gamma d - \delta}\label{Fitting scaling},
\end{equation*}
also is extended to account for rotation operators, yielding the relation 

\begin{equation}
    p_{\text{log}}^R = \alpha \left( \frac{p_{\text{phy}}}{\beta} \right)^{\gamma d_R - \delta}.\label{SD scaling form}
\end{equation}
Where \( \alpha \) is a fitting parameter, \( \beta \) serves as an error-rescaling factor, \( \gamma \) and \( \delta \) regulate the threshold scaling behavior, which depend on SD and SI  noise model respectively as given below:

\begin{equation}
    \begin{aligned}
         &\gamma = \frac{1}{2} + \frac{1}{2d e^{-\lambda (\theta^2 + \phi^2)}}, \quad \delta = \frac{1}{2} - \frac{d e^{-\lambda (\theta^2 + \phi^2)}}{2}.\\
          &\gamma = \frac{1}{2} + \frac{1}{d e^{-\lambda (\theta^2 + \phi^2)}},\quad \delta = 1- \frac{d e^{-\lambda (\theta^2 + \phi^2)}}{2}. \label{parameters}
    \end{aligned}
\end{equation}
These approaches explicitly integrates the effect of \(d_R\), showing how the \(p_{\text{log}}^R\) scales with physical errors under rotation-induced deformations in both noise models differently.

\section{Results and Discussion}\label{sec:III}

This section analyzes how rotation operators affect the code distance, revealing its exponential decay with increasing rotation angle and \(\lambda\). The logical error rate \( p_{\text{log}}^R \) decreases more rapidly with \( p_{\text{phy}} \) for small rotation angles. In quantum circuit, error channels  \(\mathcal{E}_{SD}(\theta, \phi)\) and \(\mathcal{E}_{SI}(\theta, \phi)\) define rotation-dependent noise scaling, directly impacting \( d_R \). A scaling fitting function compares logical error rates under SD and SI noise models, showing a faster decline for SI. This analysis captures \( p_{\text{log}}^R \) decay with \( d_R \) across various physical error rates. Threshold calculations and scaling analyses highlight the improved error suppression of rotation-based error correction over traditional stabilizer codes.

\subsection{Effect of rotation operator to code distance}
\begin{figure}[!ht]
    \centering
    \includegraphics[width=.8\linewidth]{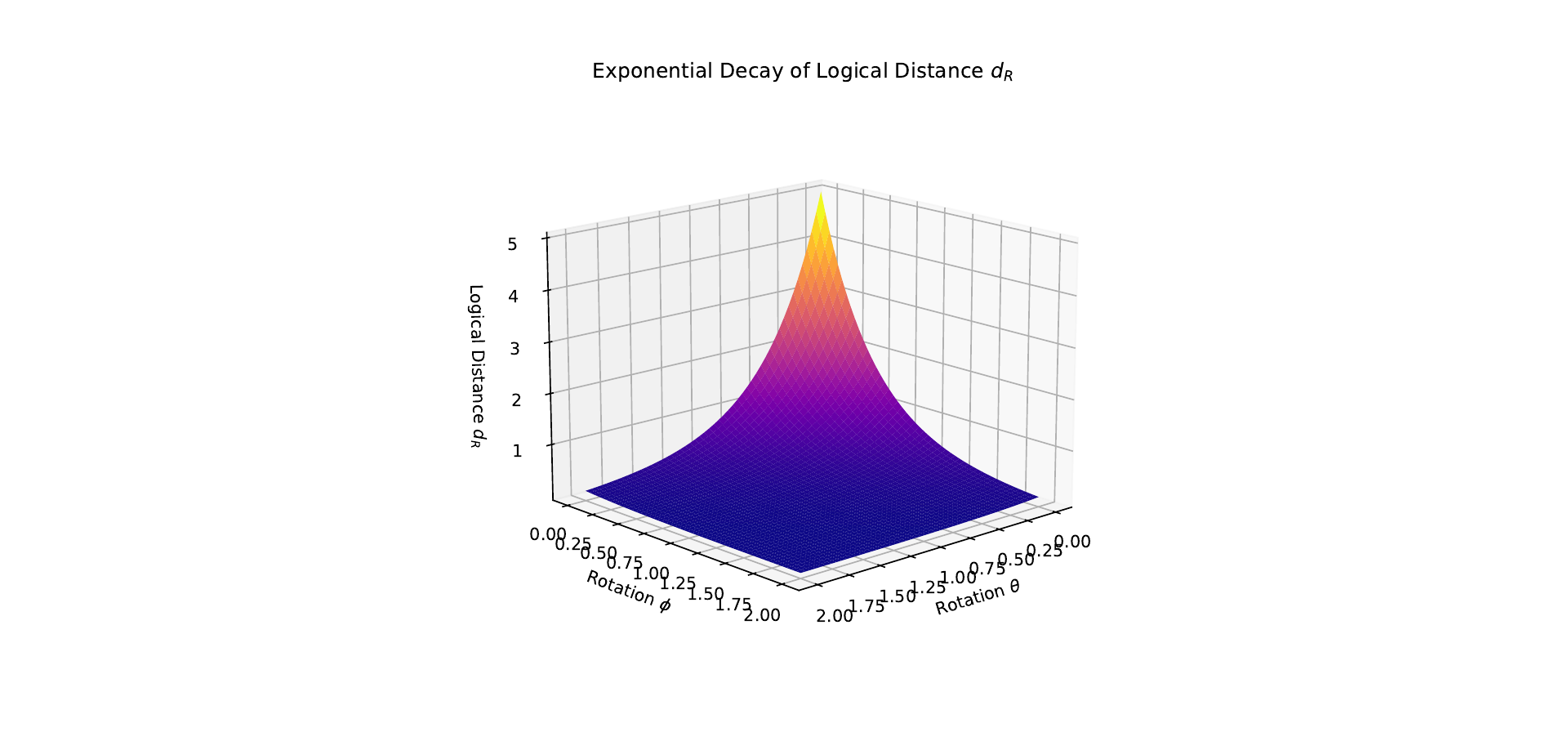}
    \caption{Effect of rotation operators on code distance  through the equation~\eqref{dR} decay exponentially to zero, as the rotation change increasingly  from \(\theta=\phi=0.1\)  to \(\theta=\phi=1.5\) at \(\lambda=1\), and \(d=5\)}
    \label{fig:effect on d}
\end{figure}

\begin{figure}[!ht]
    \centering
    \subfloat[\(\lambda=0.1\)]{\includegraphics[width=0.3\textwidth]{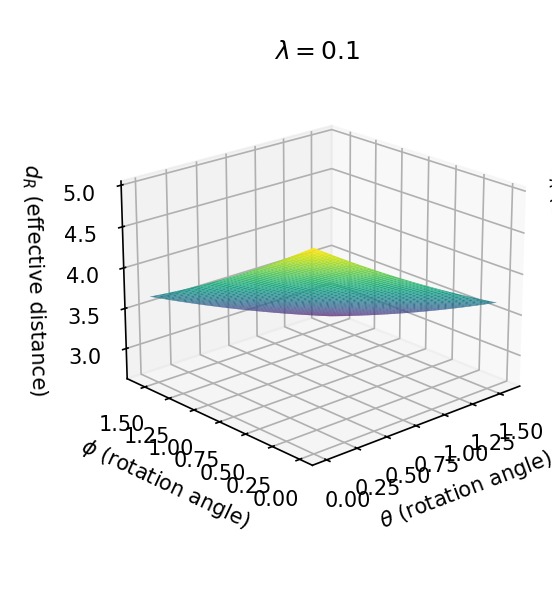}}
    \hfill
    \subfloat[\(\lambda=0.5\)]{\includegraphics[width=0.3\textwidth]{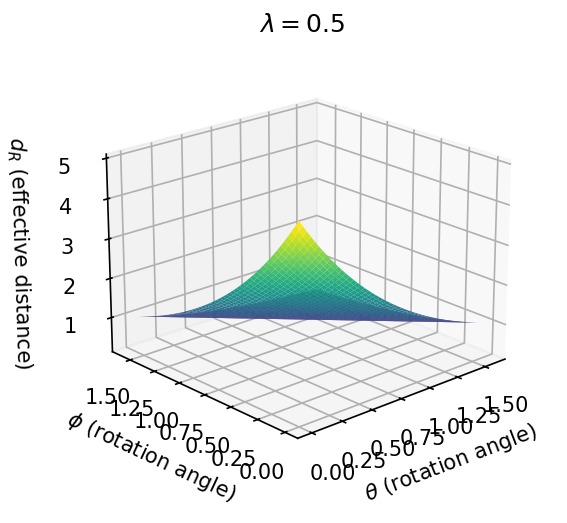}}
    \hfill
    \subfloat[\(\lambda=1.0\)]{\includegraphics[width=0.3\textwidth]{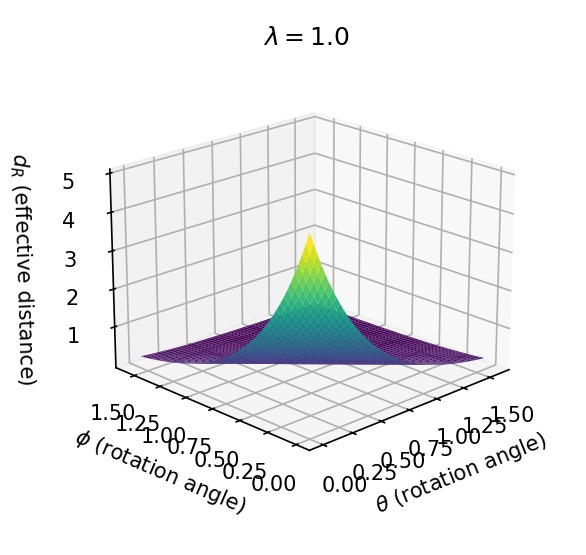}}
    \caption{ The effective code distance decay exponentially as the \(\lambda\)}.
    \label{fig:three_subfigures}
\end{figure}

 Fig.~\ref{fig:effect on d} visualizes the effect of rotation operators on code parameters by focusing on the exponential decay of the code distance as the rotation angle increases in ~\eqref{dR}. It illustrates how \(d_R\)  exponentially decreases as the rotation angle parameters \(\theta\) and \(\phi\) increase. In Fig.~\ref{fig:three_subfigures}, we show how \(d_R\) decays gradually as the value of \(\lambda\) increases, such as \(\lambda=0.1, 0.5, 1\). For small \(\lambda\) and rotation angles, \(d_R\) remains close to \(d\), which means that the error correction code is still effective. However, at large \(\lambda\) and rotation angle, \(d_R\) rapidly approaches zero, which means that the code is almost entirely ineffective in protecting against stabilizer errors. This indicates that increasing the rotation angles degrades the code distance, which shows a significant weakening of the code stabilizer error correction ability, but increasing the ability to handle error correction beyond stabilizer based. 

\begin{figure}[!ht]
    \centering
    \begin{subfigure}{0.48\textwidth}
        \centering
        \includegraphics[width=\linewidth]{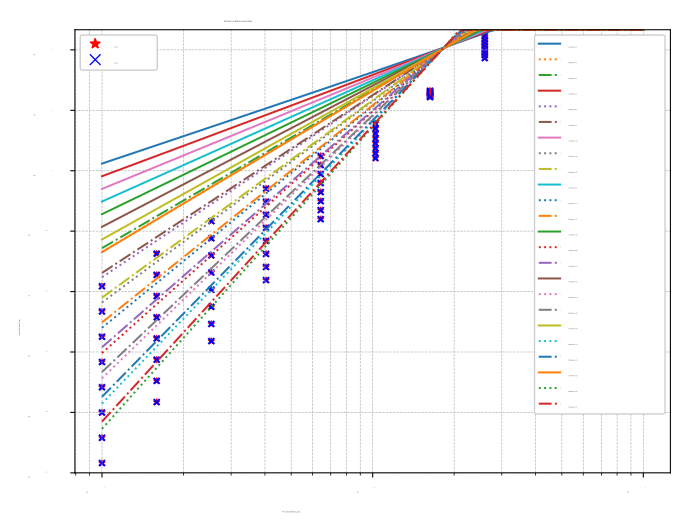} 
        \caption{ \(p_{\text{log}}^R\) as a function of \(p_{\text{phy}}\) for the SD model and rotation-induced errors, analyzed for different code distances \((d)\). The impact of small and large rotation angles is compared against the SD model.}
        \label{fig:SI and SD1}
    \end{subfigure}
    \hfill
    \begin{subfigure}{0.48\textwidth}
        \centering
        \includegraphics[width=\linewidth]{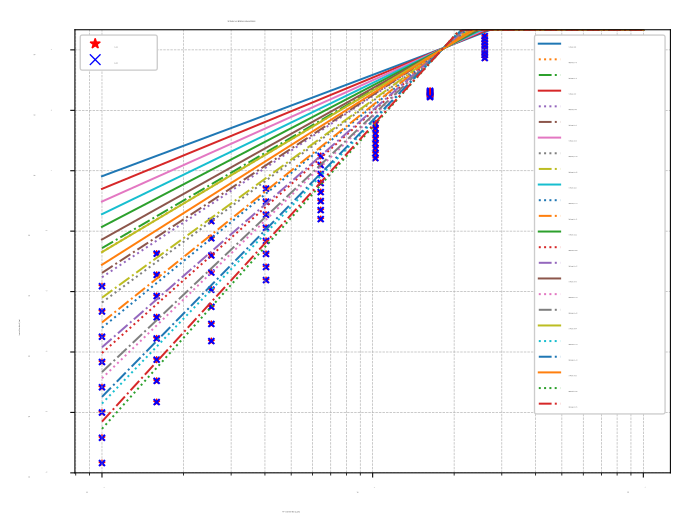} 
        \caption{\(p_{\text{log}}^R\) versus \(p_{\text{phy}}\) for the SI noise model, highlighting the influence of rotation-induced errors. Different code distances \((d)\) are examined, comparing the effects of small and large rotation angles with the SI model. }
        \label{fig:Small and larger}
    \end{subfigure}
    
    \caption{\(p_{\text{log}}^R\) as a function of the \(p_{\text{phy}}\) are derived from the universal scaling function in Eq.~\eqref{equation.Plog}, for range code distance [8-16] with the X and Z preserves the logical basis state in~\eqref{non-stab1} and the threshold is the \(p_{\text{log}}^R\) value where the curves intersect , where both has same values.}
    \label{fig:main_figure for P_log vsP_phy}
\end{figure}

Fig.~\ref{fig:main_figure for P_log vsP_phy} illustrates that the rotation operators influence \(p_{\text{log}}^R\), particularly as the code distance \((d)\) increases. The rotated logical states are affected by rotations in the Pauli X and Z basis for both small and larger rotation angle, with SI noise model exhibits a bias towards phase-flip ($Z$) errors, aligning to the superconducting qubit noise characteristics. In both the SD in Fig.~\ref{fig:SI and SD1} and the SI in Figure~\ref{fig:Small and larger} noise models, the inclusion of rotation errors results in a steeper decrease in \(p_{\text{log}}^R\) compared to standard stabilizer codes, indicating improved error suppression at higher distances. This effect is more prominent for small rotation angles, where the \(p_{\text{log}}^R\)drops faster than for larger rotation angles. The dependency of \(p_{\text{log}}^R\) on the rotation and decoherence parameters suggests that rotation-induced noise scales differently from SD and SI errors, affecting the threshold behavior of the quantum code as \(p_{th}=0.018\) and \(p_{\text{th}}=0.015\) respectively. When comparing the SI and SD models, the \(p_{\text{log}}^R\) trends exhibit a similar qualitative behavior, although the SI model generally shows a slightly lower \(p_{\text{log}}^R\) for the same  \(p_{\text{phy}}\), suggesting that the SI noise has a less detrimental effect than the SD noise. Rotation-induced errors modify these trends further, with small rotation angles maintaining a more stable error suppression, whereas larger rotations degrade performance due to accumulated phase noise. \(p_{\text{log}}^R\) drops faster toward \(10^{-22}\) due to the effect of rotation angle, which is capable to address errors beyond stabilizer-based compared to \cite{o2024compare}, demonstrating the major improvement in error correction performance.

The quantum circuits of rotated logical state~\eqref{new state} designed, for instance \(n=7\) incorporate non-Clifford gates \(\mathfrak{G}\) in~\eqref{C_ext}, 
along with rotation operators \(R_x(\theta)\) and \(R_z(\phi)\). These gates interact with logical errors through the transformed logical operators \(X_L^R\) and \(Z_L^R\) in~\eqref{logical operators}. 
The presence of noise, modeled by the SD and SI noise models, impacts the \(p_{\text{log}}^R\) differently. The Steane code, a \([[7,1,3]]\) QECCs~\cite{steane1996error,steane1998introduction}, exhibits improved error suppression when subjected to rotation operators, effectively increasing the code distance. In the SD model, the probability of logical error is approximately \(p_{\text{log,SD}} \approx 1.81 \times 10^{-3}\), while in the SI noise model, it is about \(p_{\text{log,SI}} \approx 6.62 \times 10^{-4}\). These improvements are significant compared to \(p_{\text{log}}^R\) observed without the application of rotation operators .

\begin{figure}[!ht]
    \centering
    \begin{subfigure}[b]{0.8\textwidth}
        \[
        \scalebox{0.75}{ 
        \Qcircuit @C=0.5em @R=0.5em {
            & \lstick{|\psi_R\rangle_0} & \gate{R_x(\theta)} & \gate{R_z(\phi)} & \gate{CH} & \gate{T} & \gate{CS} & \ctrl{1} & \gate{X} &\gate{Z}& \gate{\mathcal{E}_{SD}(\theta, \phi)} & \gate{R_z(-\phi)} & \gate{R_x(-\theta)} & \qw \\
            & \lstick{|\psi_R\rangle_1} & \gate{R_x(\theta)} & \gate{R_z(\phi)} & \gate{CH} & \gate{T} & \gate{CS} & \targ & \ctrl{1} & \gate{\mathcal{E}_{SD}(\theta, \phi)} & \gate{X} &\gate{Z}& \gate{R_z(-\phi)} & \gate{R_x(-\theta)} & \qw \\
            & \lstick{|\psi_R\rangle_2} & \gate{R_x(\theta)} & \gate{R_z(\phi)} & \gate{CH} & \gate{T} & \gate{CS} & \qw & \targ & \ctrl{1} & \gate{\mathcal{E}_{SD}(\theta, \phi)} & \gate{X} &\gate{Z}& \gate{R_z(-\phi)} & \gate{R_x(-\theta)} & \qw \\
            & \lstick{|\psi_R\rangle_3} & \gate{R_x(\theta)} & \gate{R_z(\phi)} & \gate{CH} & \gate{T} & \gate{CS} & \qw & \qw & \targ & \ctrl{1} & \gate{\mathcal{E}_{SD}(\theta, \phi)} & \gate{X} &\gate{Z}& \gate{R_z(-\phi)} & \gate{R_x(-\theta)} & \qw \\
            & \lstick{|\psi_R\rangle_4} & \gate{R_x(\theta)} & \gate{R_z(\phi)} & \gate{CH} & \gate{T} & \gate{CS} & \qw & \qw & \qw & \targ & \gate{\mathcal{E}_{SD}(\theta, \phi)} & \gate{X} &\gate{Z}& \gate{R_z(-\phi)} & \gate{R_x(-\theta)} & \qw \\
            & \lstick{|\psi_R\rangle_5} & \gate{R_x(\theta)} & \gate{R_z(\phi)} & \gate{CH} & \gate{T} & \gate{CS} & \qw & \qw & \qw & \qw & \targ & \gate{\mathcal{E}_{SD}(\theta, \phi)} & \gate{X} &\gate{Z}& \gate{R_z(-\phi)} & \gate{R_x(-\theta)} & \qw \\
            & \lstick{|\psi_R\rangle_6} & \gate{R_x(\theta)} & \gate{R_z(\phi)} & \gate{CH} & \gate{T} & \gate{CS} & \qw & \qw & \qw & \qw & \qw & \targ & \gate{\mathcal{E}_{SD}(\theta, \phi)} & \gate{X} &\gate{Z}& \gate{R_z(-\phi)} & \gate{R_x(-\theta)} & \qw
            }  }\]
        \caption{Effect of SD Noise (\( n=7 \)) with rotation-dependent error scaling} \label{SD1}
    \end{subfigure}
    
    \hfill
    \begin{subfigure}[b]{1\textwidth}
        \[
        \scalebox{0.75}{ 
        \Qcircuit @C=0.5em @R=0.5em {
            & \lstick{|\psi_R\rangle_0} & \gate{R_x(\theta)} & \gate{R_z(\phi)} & \gate{CH} & \gate{T} & \gate{CS} & \ctrl{1} & \gate{X} &\gate{Z}& \gate{\mathcal{E}_{SI}(\theta, \phi)} & \gate{R_z(-\phi)} & \gate{R_x(-\theta)} & \qw \\
            & \lstick{|\psi_R\rangle_1} & \gate{R_x(\theta)} & \gate{R_z(\phi)} & \gate{CH} & \gate{T} & \gate{CS} & \targ & \ctrl{1} & \gate{\mathcal{E}_{SI}(\theta, \phi)} & \gate{X} &\gate{Z}& \gate{R_z(-\phi)} & \gate{R_x(-\theta)} & \qw \\
            & \lstick{|\psi_R\rangle_2} & \gate{R_x(\theta)} & \gate{R_z(\phi)} & \gate{CH} & \gate{T} & \gate{CS} & \qw & \targ & \ctrl{1} & \gate{\mathcal{E}_{SI}(\theta, \phi)} & \gate{X} &\gate{Z}& \gate{R_z(-\phi)} & \gate{R_x(-\theta)} & \qw \\
            & \lstick{|\psi_R\rangle_3} & \gate{R_x(\theta)} & \gate{R_z(\phi)} & \gate{CH} & \gate{T} & \gate{CS} & \qw & \qw & \targ & \ctrl{1} & \gate{\mathcal{E}_{SI}(\theta, \phi)} & \gate{X} &\gate{Z}& \gate{R_z(-\phi)} & \gate{R_x(-\theta)} & \qw \\
            & \lstick{|\psi_R\rangle_4} & \gate{R_x(\theta)} & \gate{R_z(\phi)} & \gate{CH} & \gate{T} & \gate{CS} & \qw & \qw & \qw & \targ & \gate{\mathcal{E}_{SI}(\theta, \phi)} & \gate{X} &\gate{Z}& \gate{R_z(-\phi)} & \gate{R_x(-\theta)} & \qw \\
            & \lstick{|\psi_R\rangle_5} & \gate{R_x(\theta)} & \gate{R_z(\phi)} & \gate{CH} & \gate{T} & \gate{CS} & \qw & \qw & \qw & \qw & \targ & \gate{\mathcal{E}_{SI}(\theta, \phi)} & \gate{X} &\gate{Z}& \gate{R_z(-\phi)} & \gate{R_x(-\theta)} & \qw \\
            & \lstick{|\psi_R\rangle_6} & \gate{R_x(\theta)} & \gate{R_z(\phi)} & \gate{CH} & \gate{T} & \gate{CS} & \qw & \qw & \qw & \qw & \qw & \targ & \gate{\mathcal{E}_{SI}(\theta, \phi)} & \gate{X} &\gate{Z}& \gate{R_z(-\phi)} & \gate{R_x(-\theta)} & \qw
        } } \]
        
        \caption{Effect of SI Noise (\( n=7 \)) with rotation-dependent error scaling}\label{SI1}
    \end{subfigure}
    \caption{Quantum circuits of rotated logical state \eqref{new state} with effective code distance scaling (\( d_R = d e^{-\lambda (\theta^2 + \phi^2)} \)) under SD and SI noise models,  where \( \mathcal{E}_{SD}(\theta,\phi) \) and \( \mathcal{E}_{SI}(\theta,\phi) \) represent the respective noise effects with rotational angles for \( n=7 \). The non-Clifford gate~\eqref{non-Clifford gate},rotation gate~\eqref{rotation operators}, and logical gate~\eqref{logical operators} are applied.}
    \label{fig:qcurcuit}
\end{figure}
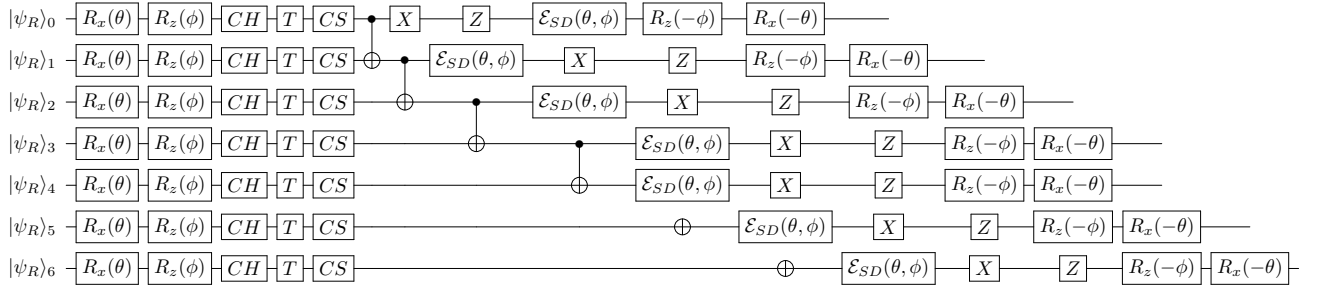
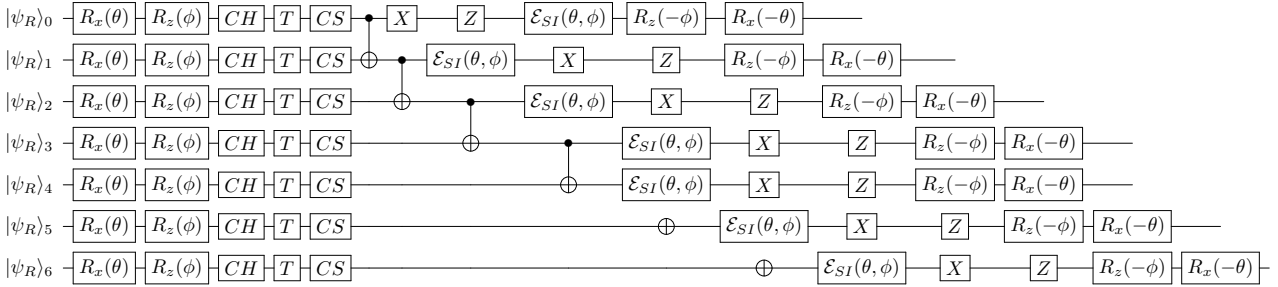

\noindent Fig.~\ref{fig:qcurcuit} presents quantum circuits of seven-qubit system, which are encoded rotated logical qubit, under the effects of rotation on the SD and SI noise models. These circuits integrate single-qubit rotations (\(R_x(\theta), R_z(\phi)\)), non-Clifford gate operations (\(CH, T, CS\)), and controlled CNOT interactions to simulate logical encoding and noise propagation.  The error channels (\(\mathcal{E}_{SD}(\theta, \phi)\))  and \(\mathcal{E}_{SI}(\theta, \phi)\)) model rotation-dependent noise scaling, influencing the effective code distance. Fig.~\ref{SD1} illustrates quantum circuit on the SD noise model, errors occur uniformly as random operators $X$, $Z$ on each qubit, making it challenging to isolate specific error types. Quantum circuit implements a sequence of rotations \(R_x(\theta)\) and \(R_z(\phi)\) to initialize each qubit, followed by a series of non-Clifford gates, which introduce quantum correlations and non-linearity essential for universal quantum computation. The presence of the Toffoli  and T-gates further entangles the qubits, potentially encoding rotated logical qubits. Standard depolarization noise \(\mathcal{E}_{SD}(\theta, \phi)\) is applied, simulating realistic quantum errors that degrade coherence under the effect of rotation. Correction or mitigation of errors is suggested by applying the corrective gates $X$ and $Z$, followed by inverse rotations \(R_z(-\phi)\) and \(R_x(-\theta)\), which may attempt to restore the original quantum state. 
Fig.~\ref{SI1} reveals that the quantum circuit structure remains largely unchanged, except that the noise model is replaced with \(\mathcal{E}_{SI}(\theta, \phi)\))), which represents a rotation-dependent probabilistic error model rather than the standard depolarization model. This implies that the noise introduced here scales differently, possibly simulating systematic errors due to imperfect rotation gate implementations rather than purely random depolarization. The influence of this noise type on QEC schemes could be different, as systematic errors tend to accumulate rather than average out as in depolarization.  The SD noise model assumes uniform depolarization across, qubits and captures bias-dependent phase errors relevant to superconducting qubits. 

\subsection{Scaling of Logical to Physical Error Rates with Rotation Effect}

Threshold values for the SD and SI noise models at small and large rotation angles, with a scaling exponent \(\nu_0=1.01\), are given in \eqref{thre-lo-phy} and \eqref{eq:thres SI}. For SD, \( p_{th} = 0.018 \), while for SI, \( p_{th} = 0.015 \), as shown in Fig.~\ref{fig:main_figure for P_log vsP_phy}. These thresholds are comparable to the honeycomb code results in \cite{gidney2021fault} for two-body measurements and significantly align with \cite{stephens2014fault}, which modifies the standard noise model with perfect single-qubit gates. Although lower than the values in \cite{wang2003confinement} for the random-plaquettes gauge model in even rounds, they exceed the thresholds of \(\approx 0.0124\) and \(\approx 0.00995\) for SD and SI, respectively, reported in \cite{o2024compare}.

\vskip3mm

\begin{table}[!ht]
    \centering
    \caption{The combined fitting of odd and even code distances determines the parameters for the SD and SI noise models under the influence of rotation angles, following in~\eqref{SD scaling form}, as discussed in ~\cite{o2024compare}. The parameter values are presented in Table~\ref{tab:noise_model}.}
    \vskip2mm
    
    \begin{tabular}{llc} \hline \textbf{Noise Model} & \textbf{Rotation Type} & \textbf{Equation} \\ \hline \multirow{2}{*}{SD} & Small Angle & \( p_{\text{log}}^R = 0.065 \left( \frac{p_{phy}}{0.0044} \right)^{0.68 d_R +0.87}\) \\ 
    & Large Angle & \( p_{\text{log}}^R = 0.063 \left( \frac{p_{phy}}{0.0064} \right)^{0.65 d_R +1.12}\) \\ \hline 
    \multirow{2}{*}{SI} & Small Angle & \( p_{\text{log}}^R = 0.064 \left( \frac{p_{phy}}{0.0042} \right)^{0.81 d_R +0.62}\) \\ 
    & Large Angle & \( p_{\text{log}}^R = 0.034 \left( \frac{p_{phy}}{0.0057} \right)^{0.77 d_R +0.87}\) \\ \hline 
    \end{tabular}
    
    \label{tab:noise_model23}
\end{table}

\begin{figure}[!ht]
    \centering
    \includegraphics[width=.8\linewidth]{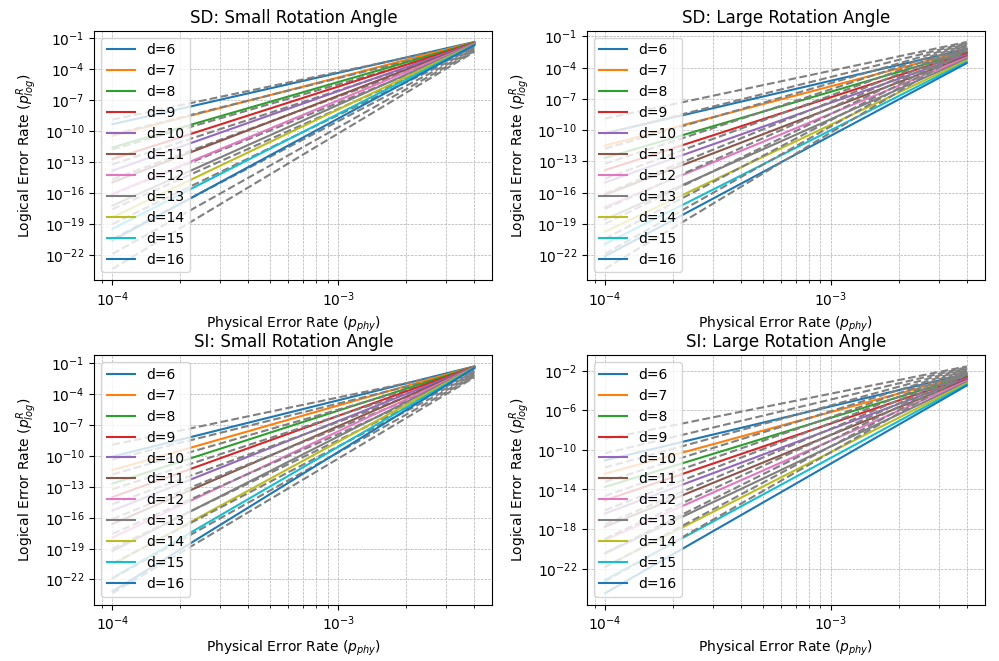}
    \caption{Function fit plots displaying the fitted function (Eq.~\eqref{SD scaling form}) as grey dashed lines, with Equations presented in Table~\ref{tab:noise_model23}. This results are close to the Fig.~\ref{fig:main_figure for P_log vsP_phy} but differ on \(p_{\text{phy}}\) decreased up to \(10^{-4}\) }
    \label{fig:enter-label}
\end{figure}

The fitting scaling function in Table~\ref{tab:noise_model23} presents the \( p_{\text{log}} \) formula with \( d_R \), integrating parameter values from Table~\ref{tab:noise_model} into~\eqref{SD scaling form}. SI noise exhibits slightly better scaling than SD noise. Fig.~\ref{fig:enter-label} presents fitted curves over sampled data, with parameter values for odd, even, and combined code distances listed in Table~\ref{tab:noise_model}. The logical error rate \( p_{\text{log}}^R \) decreases to \( 10^{-22} \) for both SD and SI, surpassing \cite{o2024compare} and significantly improves over standard stabilizer QEC, effectively addressing coherent errors challenges in near-term quantum devices. Line fits of \( p_{\text{log}}^R \) versus \( p_{\text{phy}} \) curves for \( p_{\text{phy}} \leq 10^{-4} \) are performed fro each code distance. While logical qubit rotations preserve qubit count, they modify error correction by reducing the code distance and error threshold compared to ~\cite{wang2003confinement}. These effects vary by noise model, with SD experiencing increased non-Pauli mixing and SI losing its bias-preserving advantage. The fitted scaling function offers a predictive framework for \( p_{\text{log}}^R \) under rotations, aiding fault-tolerant quantum computation optimization.
\vskip3mm

\begin{table}[!ht]
    \centering
    \caption{Noise model characteristics for different rotation types, fitted to Eq.~\eqref{SD scaling form} using \( d \geq 2 \) and \( p_{phy} \leq 0.004 \). The optimal \( \gamma \) and \( \delta \) were obtained from line-fit gradients in Eq.~\eqref{parameters}. The combined fit accounts for both odd and even code distances, with \( d_R \) modified by overlapping small and large rotation angle effects within uncertainty.}
\vskip3mm
    
    \begin{tabular}{lllccccc}
        \toprule
        \multirow{2}{*}{Noise Model} & \multirow{2}{*}{Rotation Type}& \multirow{2}{*}{$d$} & \multirow{2}{*}{$d_R$} & \multicolumn{4}{c}{Parameters} \\
        \cmidrule(lr){5-8}
        & & & & \(\alpha\) & \(\beta\) & \(\gamma\) & \(\delta\) \\
        \midrule
        \multirow{6}{*}{SD} & \multirow{2}{*}{Small Angle}&even&  2.7498  & 0.0644  & 0.0044  & 0.6818  & -0.8749 \\
        & & odd& 3.2497  & 0.0622  & 0.0064  & 0.6539  & -1.1249 \\
        \cmidrule(lr){2-8}
        & \multirow{2}{*}{Large Angle} &even&2.7445  & 0.0645  & 0.0062  & 0.6822  & -0.8723 \\
        & & odd& 3.2435  & 0.0978  & 0.0042  & 0.6542  & -1.1218 \\
        \cmidrule(lr){2-8}
        & \multirow{2}{*}{Combined} && 2.7400  & 0.0645  & 0.0044  & 0.6825  & -0.8700 \\
        & & & 3.2380  & 0.0623  & 0.0064  & 0.6544  & -1.1190 \\
        \midrule
        \multirow{6}{*}{SI} & \multirow{2}{*}{Small Angle} &even& 3.2497  & 0.0643  & 0.0042  & 0.8077  & -0.6249 \\
        & & odd&3.7497  & 0.0664  & 0.0039  & 0.7667  & -0.8749 \\
        \cmidrule(lr){2-8}
        & \multirow{2}{*}{Large Angle} &even& 3.2435  & 0.0357  & 0.0042  & 0.8083  & -0.6218 \\
        & & odd&3.7425  & 0.0664  & 0.0039  & 0.7672  & -0.8713 \\
        \cmidrule(lr){2-8}
        & \multirow{2}{*}{Combined} &&3.2390  & 0.0643  & 0.0042  & 0.8087  & -0.6195 \\
        & & & 3.7370  & 0.0336  & 0.0057  & 0.7676  & -0.8685 \\
        \bottomrule
    \end{tabular}
    \label{tab:noise_model}
\end{table}

\begin{figure}[!ht]
    \centering
    \includegraphics[width=.7\linewidth]{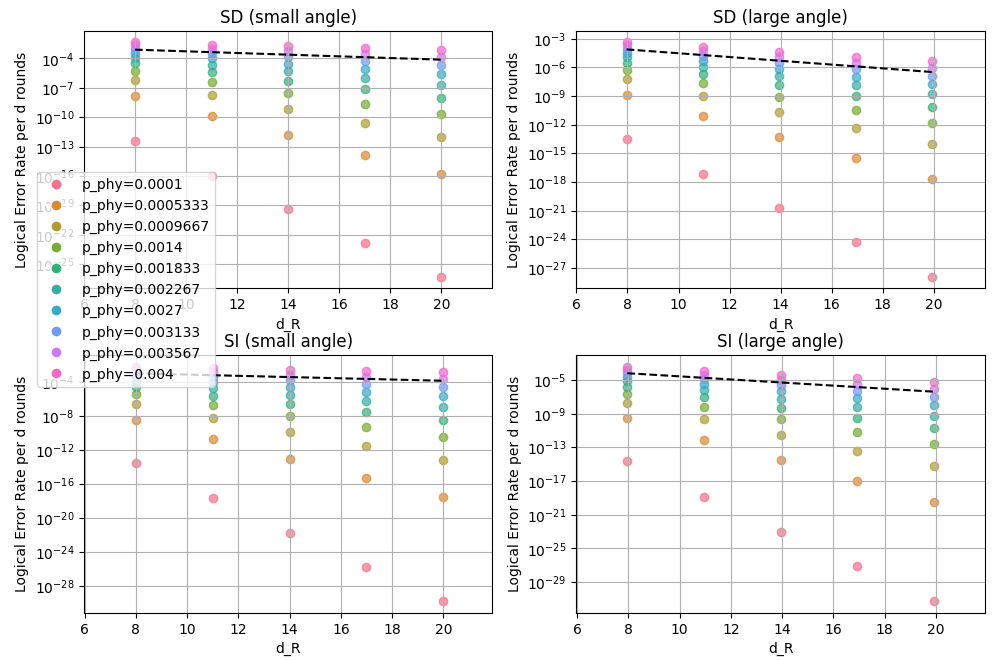}
    \caption{Relationship between \( p_{\text{log}}^R \) and \( d_R \) for SD and SI noise models from Table~\ref{tab:noise_model23}, fitted to Eqs.~\eqref{parameters} and \eqref{dR}. The dashed black curve shows the decreasing fitted scaling function for larger rotations, extending beyond QEC. A rapid drop in \( p_{\text{log}}^R \) from \( 10^{-4} \) to \( 10^{-29} \) for SI occurs as \( d_R \) increases, across \(p_{\text{phy}}\) between \( 10^{-3} \) and \( 10^{-4} \), indicating enhanced error correction. For small rotation angles, stability is maintained but weakens as the angle grows.  } 
    \label{fig:d_R vs logical}
\end{figure}

Fig.~\ref{fig:d_R vs logical} illustrates the exponential decay of \( p_{\text{log}}^R \) with \( d_R \), confirming the effectiveness of QEC. However, SD noise, associated with traditional stabilizer codes, exhibits slower error suppression and greater sensitivity to rotation angles, leading to higher \( p_{\text{log}}^R \) under large coherent errors. In contrast, SI noise, representing beyond-stabilizer codes, achieves superior error suppression, with \( p_{\text{log}}^R \) decaying up to \( \geq 10^{-29} \) and reduced sensitivity to coherent noise. Using the same approach as in Table~\ref{tab:noise_model}, we analyze \( d \) rounds between [8–23], leading to \( d_R \) values of [5.999–21.9965] for SD and [5.9760–21.9122] for SI, following Eq.~\eqref{dR}. While \( d \) decreases to \( d_R \) more slowly, the approach increases error correction beyond traditional methods. SI noise models demonstrate better error suppression, particularly at high \( p_{\text{phy}} =10^{-4} \) for both small and large rotation angles. This highlights the potential of beyond-stabilizer approaches in providing stronger protection against coherent errors, making them a promising alternative for practical quantum computing. The fitted curves (dashed lines) depict how \( p_{\text{log}}^R \) scales with \( d_R \), revealing key trends in error correction performance.

\section{Conclusion}\label{sec:IV}

This work extends the stabilizer logical state to rotated logical state under the rotation operators. Its rotated stabilizer generators are no longer a pure Pauli operator but a linear combination of multiple Pauli terms. For the rotated logical state at large rotations, stabilizers significantly mix into non-Pauli terms through the rotation operators, offering a promising alternative by utilizing non-commutative structures.
Our results establish that the effective code distance \(d_R\) is essential in logical error suppression, with an exponential decay in \(p_{\text{log}}^R\) with rotation angles. The rotation operators extend the code distance by transforming stabilizer generators into non-Pauli forms, improving error resilience beyond conventional stabilizer codes. Numerical simulations reveal that SI noise achieves stronger error suppression than SD noise, with threshold error rates ranging between \(0.018\) and \(0.015\) respectively compared to \cite{stephens2014fault,o2024compare,wang2003confinement}. At large rotation angles for SI with \(d_R\) result shows a steeper decay of \(p_{\text{log}}^R=10^{-29}\) which introduces non-Pauli errors, modifying logical error behavior, while large rotations for SD maintains high range \(10^{-3}\) to \(10^{-27}\) of \(p_{\text{log}}^R\) for \(d_R=20\) and \(p_{\text{phy}}=10^{-4}\). The graphical analysis of logical versus physical error rates highlights the superior performance of SI noise in mitigating errors. The logical error rate continue decreasing exponentially, suggesting stronger suppression of logical error beyond traditional stabilizer codes. Additionally, as rotations extend logical states to \(|0_L^R\rangle\) and \(|1_L^R\rangle\), their stabilizer-based protection remains effective for small angles but weakens for larger rotations. These findings show that rotation-based encoding strategies increase the effective code distance and improve error correction beyond standard stabilizer formalism, offering new pathways for fault-tolerant quantum computation.

\subsection*{Acknowledgment}
This work is supported by the Air Force Office of Scientific Research under Award No. FA2386-22-1-4062. We also acknowledge the use of IBM Quantum services in conducting this study.
\subsection*{ Conflict and Interest}
The authors declare that there are no conflicts of interest associated with this work.\\

\textbf{Declaration of generative AI and AI-assisted technologies in the writing process}\\
During the preparation of this work the author(s) used CHATGPT in order to improve language and readability of the work. After using this tool/service, the authors reviewed and edited the content as needed and take full responsibility for the content of the publication.

\bibliographystyle{unsrt}
\bibliography{references. bib}

\end{document}